\documentclass[aps,prb,preprint,groupedaddress,showpacs]{revtex4}
\usepackage{graphicx}
\usepackage{helvet}
\usepackage{amsmath,amssymb}
\usepackage{bm}
\usepackage{subfigure}

\makeatletter
\newcommand{\Mtmdt}{[M(tmdt)$_{2}$]}
\newcommand{\Au}{[Au(tmdt)$_{2}$]}
\newcommand{\Cu}{[Cu(tmdt)$_{2}$]}
\newcommand{\Ni}{[Ni(tmdt)$_{2}$]}
\newcommand{\proton}{$^{1}$H}
\newcommand{\muB}{${\mu}_{\rm B}$}
\newcommand{\Tn}{$T_{\rm N}$}

\newcommand{\Toner}{$T_{1}^{-1}$}
\makeatother
\begin{document}

\title{NMR Evidence for Antiferromagnetic Transition in the Single-Component Molecular System, [Cu(tmdt)$_{2}$]}
\author{Rina\ Takagi,\ Kazuya\ Miyagawa,\ and\ Kazushi\ Kanoda}
\affiliation{Department\ of\ Applied Physics,\ University\ of\ Tokyo,\ Bunkyo\ City,\ Tokyo,\ 113-8656,\ Japan}
\author{Biao\ Zhou,\ Akiko\ Kobayashi,\ and\ Hayao\ Kobayashi}
\affiliation{Department\ of\ Chemistry,\ College\ of\ Humanities\ and\ Sciences,\ Nihon\ University,\ Setagaya-ku,\ Tokyo,\ 156-8550,\ Japan} 
 
\date{\today}

\begin{abstract}
 The magnetic state of the single-component molecular compound, [Cu(tmdt)$_{2}$], is investigated by means of $^{1}$H-NMR.
 An abrupt spectral broadening below 13 K and a sharp peak in nuclear spin-lattice relaxation rate, $T_{1}^{-1}$, at 13 K are observed as clear manifestations of a second-order antiferromagnetic transition, which is consistent with the previously reported magnetic susceptibility and EPR measurement.
 The ordered moment is estimated at $0.22-0.45$ ${\mu}_{\rm B}$/molecule.
 The temperature-dependence of $T_{1}^{-1}$ above the transition temperature indicates one-dimensional spin dynamics and supports that the spins are on the central part of the molecule differently from other isostructural compounds.
\end{abstract}

\maketitle

\section{Introduction}
 Most of molecular conductors are charge transfer salts composed of donor and acceptor molecules, either or both of which are responsible for electric conduction and magnetism. The conduction band is constructed by either HOMO (highest occupied molecular orbital) or LUMO (lowest unoccupied molecular orbital) without mixture because their large energy separation.
 However, single-component molecular conductors, {\Mtmdt},\cite{1} are pointed out to have quasi-degenerate molecular orbitals by the first-principles band calculations,\cite{2} where M and tmdt stand for a metallic ion and trimethylenetetrathiafulvalenedithiolate, respectively.
 The frontier molecular orbitals are composed of the ${\pi}$-type orbitals extending over the tmdt ligands (symmetric and asymmetric $p{\pi}$ orbitals) and the ${\sigma}$-type orbital localized around the metal $d$ orbital ($dp{\sigma}$ orbital).
 Their energy separation depends on the metallic ions, M; that is, the hybridization of $p{\pi}$ and $d$ orbitals is varied by the energy level of $d$ orbital in M relative to the level of $p{\pi}$ orbital.
 Among various isostructural compounds {\Mtmdt}, the $p{\pi}$ and $dp{\sigma}$ orbitals are energetically closer in M = Au and Cu compounds than the others such as {\Ni},\cite{2, 3} where only the $p{\pi}$ orbitals in tmdt construct the conduction bands with three-dimensional Fermi surfaces confirmed by de Haas-van Alphen oscillation.\cite{4} The orbital degeneracy is possibly responsible for the unique magnetic properties of M = Au and Cu compounds as follows.

 {\Au} undergoes an antiferromagnetic phase transition at {\Tn}$ = 110$ K, as evidenced by susceptibility and NMR measurements,\cite{5, 6} in spite of the metallic behavior in resistivity down to low temperatures. The first-principles calculation suggests a possible Fermi-surface instability into a spin-density-wave (SDW) in the tmdt conduction band formed by the asymmetric $p{\pi}$ orbital and predicts an antiferromagnetic spin configuration between two tmdt ligands within a molecule.\cite{2}
 However, this SDW scenario has some difficulties in explaining experimental facts. For example, there is no change in resistivity around the antiferromagnetic transition.
 In addition, assuming that the SDW moment is on the tmdt ligand, the analysis of NMR spectra yields a sizable antiferromagnetic moment of $0.7 - 1.2$ {\muB}/tmdt,\cite{6} which is considerably larger than the value, 0.08 {\muB}/tmdt, predicted in the SDW scenario.\cite{2}
 This disagreement seems to require reconsideration for the electronic state in {\Au}, taking the multi-orbital character into account.

 Meanwhile, the recently synthesized {\Cu} is found to have a new aspect in this family of single-component molecular conductors.\cite{7} According to the first-principles calculations, the Fermi level is in the conduction band formed by the $dp{\sigma}$ orbital. However, the conductivity is semiconductive down to low temperatures and the magnetic susceptibility exhibits the Bonner-Fisher type temperature dependence with the exchange interaction $J$ of 169 K.\cite{7}
 The $dp{\sigma}$ orbital of {\Cu}, localized around the central Cu ion coordinated by four S atoms, is in contrast to the $p{\pi}$ orbital extended over the tmdt ligands, and therefore the on-site Coulomb energy should be more enhanced in {\Cu} than the other {\Mtmdt} salts with the $p{\pi}$ orbital conduction-bands. Thus, the insulating state in {\Cu} is reasonably attributed to strong electron correlations.
 The intermolecular overlap integral of the $dp{\sigma}$ orbital along $a$ axis is one order of magnitude larger than those in the other two directions.\cite{7}
 Thus, if spins are on the $dp{\sigma}$ orbital, they should form quasi-one-dimensional $S=1/2$ antiferromagnetic Heisenberg chains. A characteristic field-dependence of magnetic susceptibility and a sudden decrease in EPR signal intensity suggest an antiferromagnetic phase transition at $12-13$ K.\cite{7}

\begin{figure}
\begin{center}
\includegraphics[width=8.6cm,keepaspectratio]{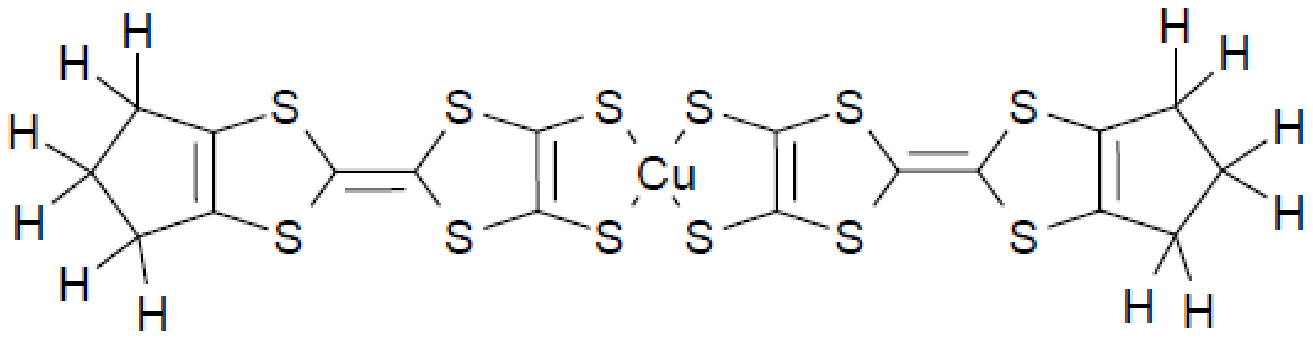}
\end{center}
\caption{Schematic of {\Cu}.}
\label{Fig1}
\end{figure}

 To characterize the magnetic state of {\Cu} from the microscopic point of view, we have carried out NMR measurements at {\proton} sites in the planar Cu(tmdt)$_{2}$ molecule shown in Fig.~\ref{Fig1}.
 Here, we report on NMR evidence for a second-order transition to an antiferromagnetic state at 13 K with a moment of $0.22-0.45$ {\muB}/molecule and spin dynamics suggesting the one-dimensional Heisenberg nature, which is consistent with the picture that the spins are on the $dp{\sigma}$ orbital in {\Cu}.

\section{EXPERIMENTAL}
 {\proton}-NMR measurements were performed for poly crystals of {\Cu}. NMR spectra and nuclear spin-lattice relaxation rate, {\Toner}, were measured under a magnetic field of 3.66 T in a temperature range from room temperature down to 1.9 K.
 The spectra were obtained by the fast Fourier transformation of echo signals observed after the so-called solid-echo pulse sequence, $({\pi}/2)_{x}-({\pi}/2)_{y}$.\cite{8} The typical pulse width was 1.2 ${\mu}$s. The spectra were much broadened below 13 K, where we set the pulse width less than 1.0 ${\mu}$s in order to cover the whole spectral frequency.
 By examining the spectra with varying the radio frequency under the fixed magnetic field, we confirmed that the present pulse condition is sufficient for getting the whole spectra properly even below 13 K.
 {\Toner} is obtained by the standard saturation-recovery method. The relaxation curves of nuclear magnetization deviated from the single exponential function by the reasons described later;
 so we define {\Toner} by fitting them to the so-called stretched-exponential function, $M({\infty})-M(t){\propto} \exp[-(t/T_{1})^{\beta}]$ in the whole temperature region. 

\section{RESULTS AND DISCUSSIONS}
\begin{figure}
\begin{center}
\includegraphics[width=8.6cm,keepaspectratio]{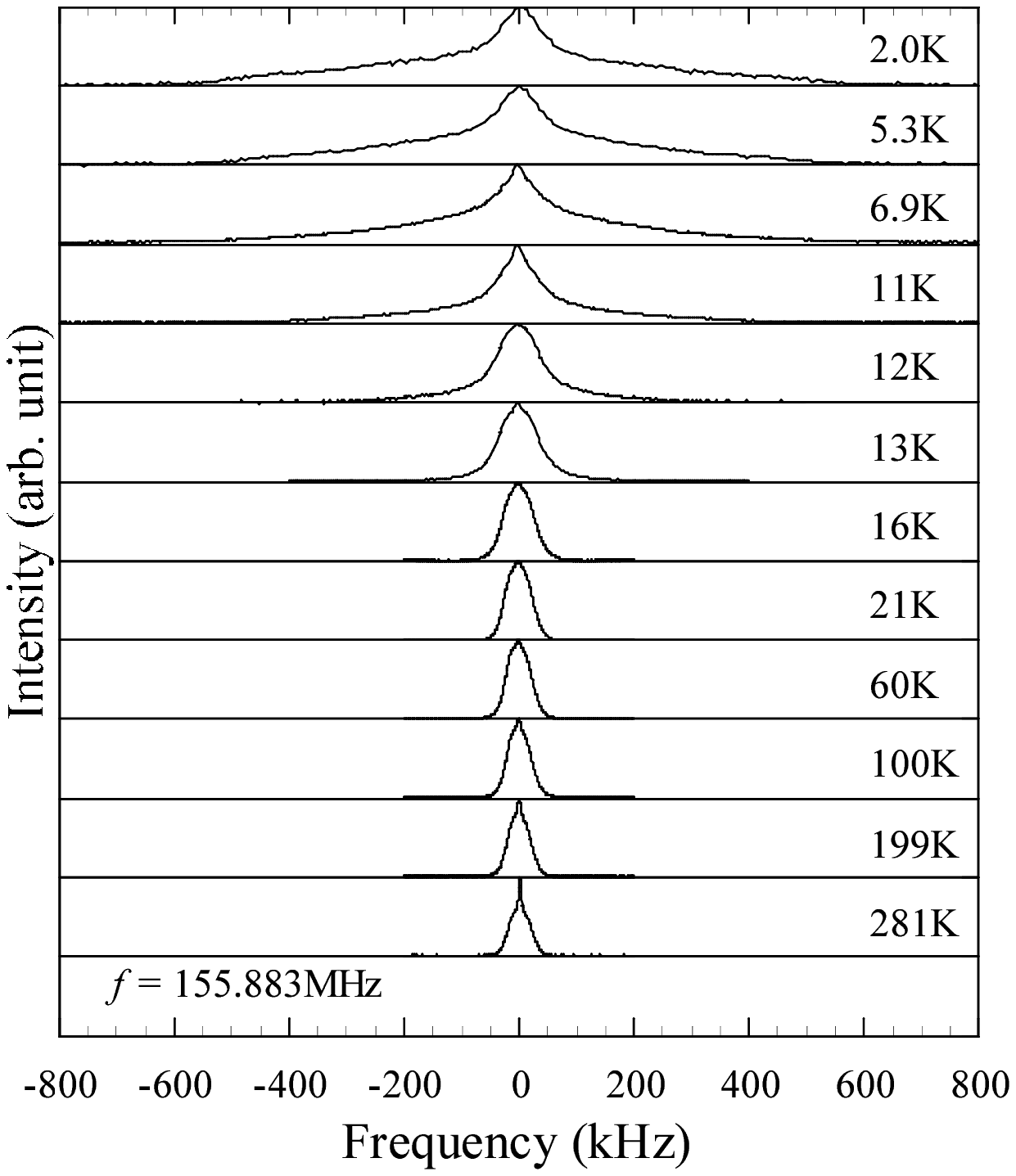}
\end{center}
\caption{Temperature dependence of {\proton}-NMR spectra of polycrystalline {\Cu}.}
\label{Fig2}
\end{figure}

\begin{figure}
\begin{center}
\includegraphics[width=8.6cm,keepaspectratio]{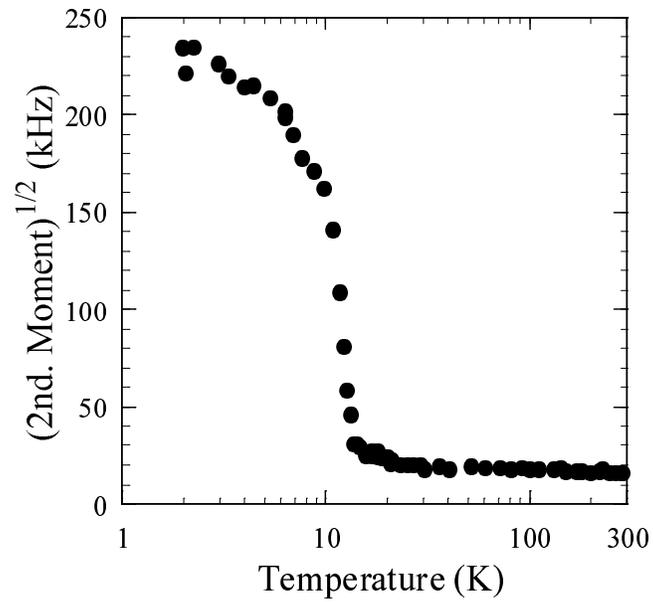}
\end{center}
\caption{Square root of the second moment as a function of temperature.}
\label{Fig3}
\end{figure}

 The {\proton}-NMR spectra are shown in Fig.~\ref{Fig2}. The spectral width characterized by the square root of second moment is about 17 kHz at room temperature (see Fig.~\ref{Fig3}), which is reasonably explained by nuclear-dipole interactions in a trimethylene group.
 Because of the small hyperfine coupling with conduction electrons, the Knight shift at proton sites is too small to be resolved, which explains why the spectral position is not changed against temperature variation, while the susceptibility shows clear temperature dependence.\cite{7}
 On the other hand, the spectra are broadened below 13 K and extended over a frequency range of as large as ${\pm}$700 kHz at 1.9 K. As shown in Fig.~\ref{Fig3}, the square root of the second moment exhibits an abrupt but continuous increase below 13 K, indicating an appearance of internal fields due to a magnetic order.
 The nuclear spin-lattice relaxation rate {\Toner} has a sharp peak, indicative of critical slowing down toward a spin ordering at the same temperature as shown in Fig.~\ref{Fig5}.
 These two features provide clear microscopic evidences for a second-order antiferromagnetic transition at {\Tn}$=13$ K.

\begin{figure}
\begin{center}
\includegraphics[width=17.8cm,keepaspectratio]{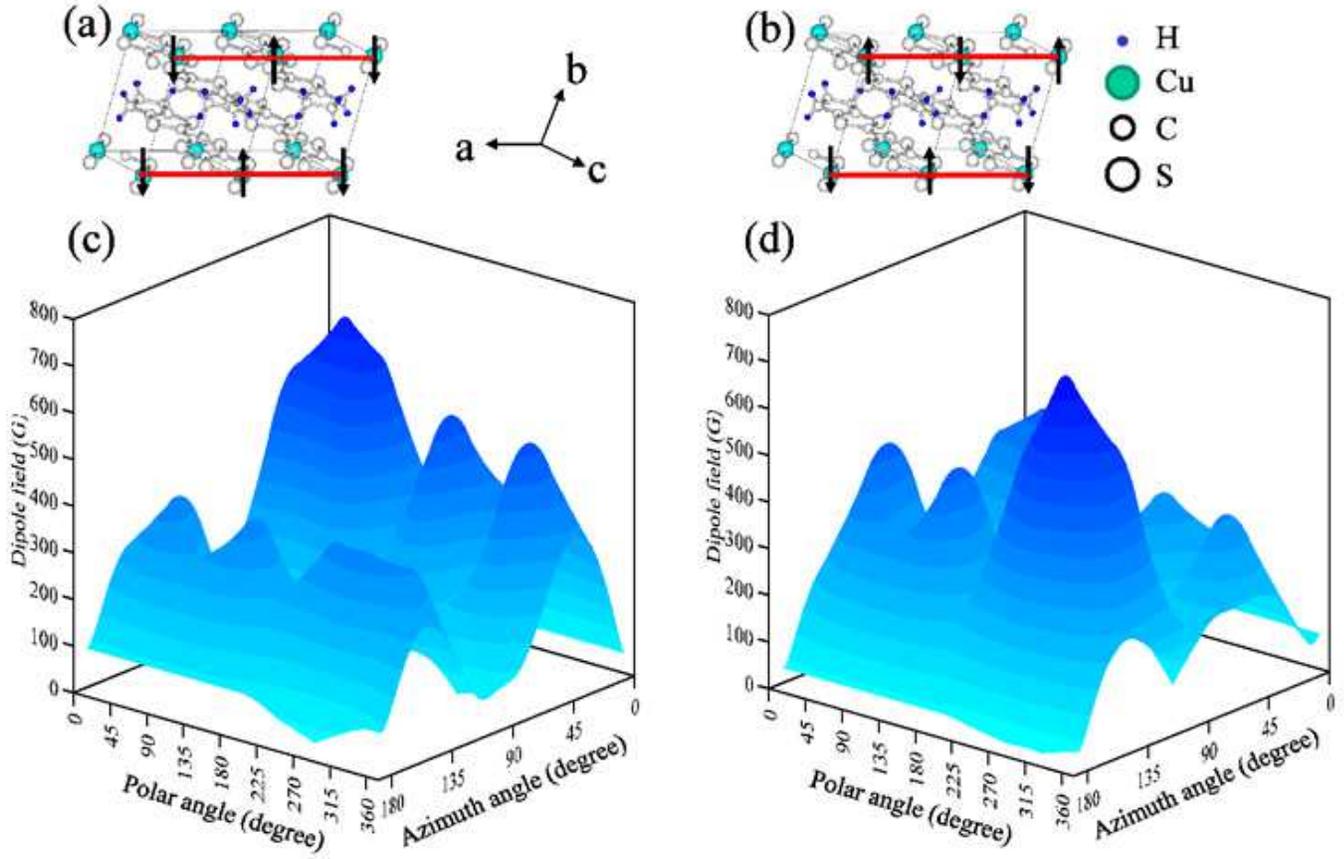}
\end{center}
\caption{(Color Online) Two possible antiferromagnetic spin configurations are shown with respect to the inter-chain coupling along $b$ axis;
 (a) ferromagnetic and (b) antiferromagnetic coupling. Antiferromagnetic chains along $a$ axis is indicated by bold solid lines.
 (c) and (d) show absolute value of dipole field parallel to the external field at one proton site, for the spin configuration patterns (a) and (b), respectively.
 These are the summation of six Cu sites, illustrated by arrows in (a) and (b).}
\label{Fig4}
\end{figure}

\begin{figure}
\begin{center}
\includegraphics[width=8.6cm,keepaspectratio]{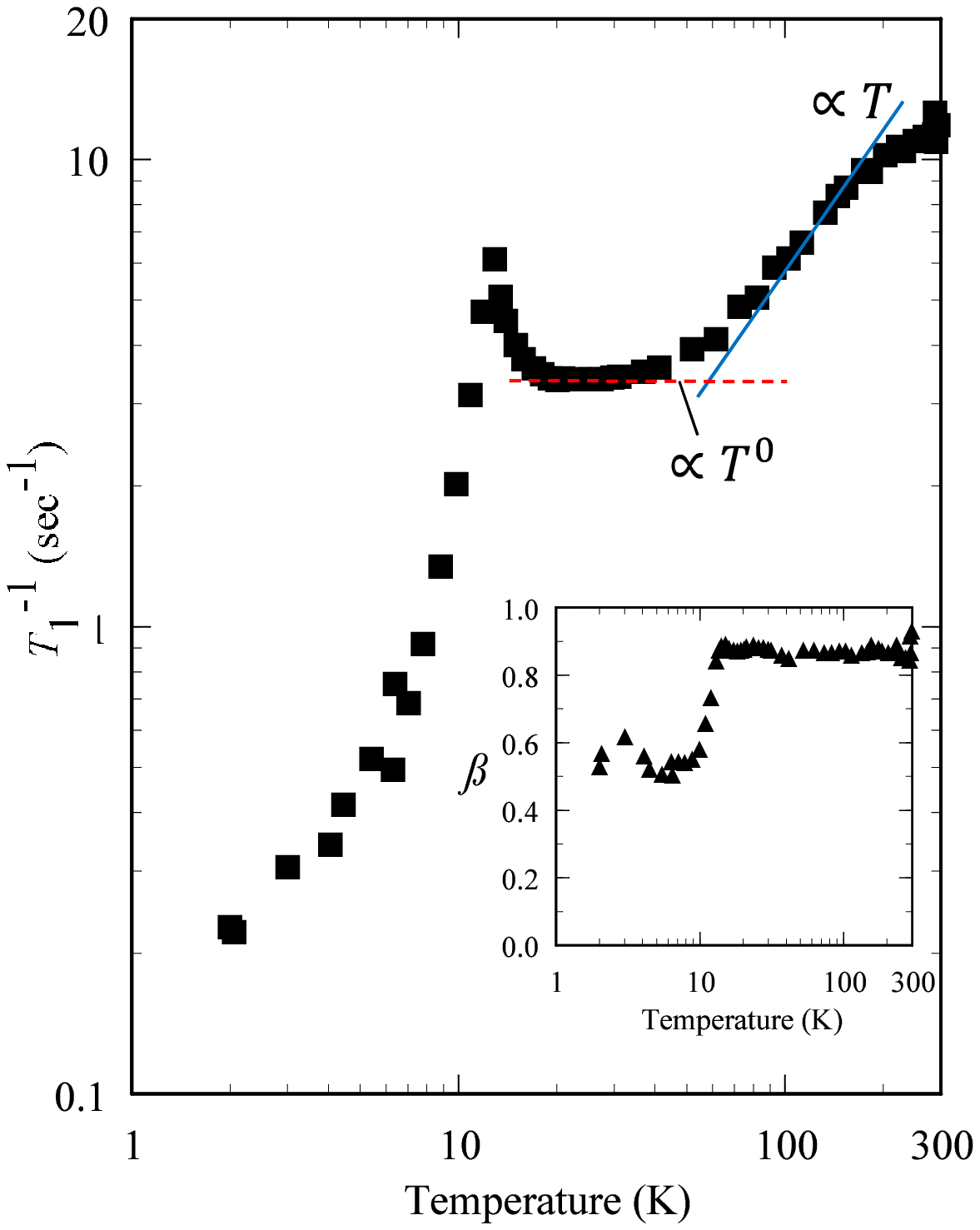}
\end{center}
\caption{(Color Online) Temperature-dependence of {\Toner} obtained by fitting to stretched-exponential functions of the relaxation curves.
 {\Toner} can be fitted to the linear function of temperature in 60 K$<T<$160 K (solid line), and temperature-independent function (constant) in 20 K$<T<$40 K (dashed line).
 The inset shows temperature-dependence of the exponent ${\beta}$, obtained as fitting parameters.}
\label{Fig5}
\end{figure}

 According to molecular-orbital calculations,\cite{7} $dp{\sigma}$ orbital is localized around the central Cu ion and the surrounding four S atoms. The tmdt ligand contains six protons available for NMR.
 The dipolar fields at each proton sites are estimated, assuming that $S=1/2$ spins are located at the central part of the molecule. Since the tmdt ligand has an elongated structure, the distance of the proton sites from Cu within a Cu(tmdt)$_{2}$ molecule is about $11.5-13.3$ {\AA}.
 It is much longer than the distance from Cu in the adjacent molecules in the crystal, which are about $2.7-4.9$ {\AA} at minimum. Since the dipolar field is proportional to ${\sim}1/r^{3}$, where $r$ is a distance between a proton and Cu atom, the main contribution to the local field at the proton sites is from the $dp{\sigma}$ spins in the six nearest-neighbor molecules [see Figs.~\ref{Fig4}(a) and~\ref{Fig4}(b)].
 In what follows, we estimate the antiferromagnetic moment from the observed spectra, using the atomic parameters obtained by x-ray diffraction study.\cite{9} As the $dp{\sigma}$ spins form a quasi-one-dimensional antiferromagnetic chain along $a$ axis, we assume an antiferromagnetic spin configuration in this direction as indicated by red lines in Figs.~\ref{Fig4}(a) and~\ref{Fig4}(b).
 With respect to the inter-chain coupling along $b$ axis, ferromagnetic or antiferromagnetic coupling is conceivable as shown in Fig.~\ref{Fig4}(a) and~\ref{Fig4}(b), respectively. Note that the spin arrangements along $c$ axis make no difference in the estimation of the local field at the proton sites.
 Assuming that a moment of 1 {\muB} is located on Cu site, we calculated the dipole fields from six Cu spins, illustrated by arrows in Figs.~\ref{Fig4}(a) and~\ref{Fig4}(b), and then summed them up to get the total local field at each proton site. In the present experiment with a powdered sample, the direction of the external magnetic field against crystal axes is different from grain to grain in the sample.
 In addition, the present field of 3.66 T is higher than the spin-flop field but the easy-axis is unknown.\cite{5} Taking these into consideration, we set the moment direction arbitrarily with keeping the spin configuration as mentioned above, and calculated the component perpendicular to the local moment aligned normal to the external field due to the spin flop.
 Note that NMR line measures local field parallel to the external field and perpendicular to the antiferromagnetic moment. The calculation was performed for all the moment direction at the six proton sites. Because the distance and direction from the nearest Cu spin are dependent on the proton sites in a tmdt, the local field is quite different from proton site to site.
 However, what we need for the estimation of an antiferromagnetic moment is only the maximum value of the local field in six protons, which corresponds to the spectral edge.
 As an example, the calculated dipole field at one proton site is shown in Figs.~\ref{Fig4}(c) and ~\ref{Fig4}(d), for the spin configuration patterns (a) and (b), respectively. The maximum value is about 680 G in pattern (a) and 740 G in pattern (b). The observed spectral edge of 700 kHz at 1.9 K, which is equivalent to 160 G in local field, points to the local moments of 0.24 {\muB}/molecule and 0.22 {\muB}/molecule respectively.
 We also performed the similar calculations, assuming that spins are distributed only on four S atoms around Cu ions, which gave, as the maximum value of the local field, 420 G in pattern (a) and 360 G in pattern (b). These values correspond to the local moments of 0.40 {\muB}/molecule and 0.45 {\muB}/molecule respectively.
 Because the real $dp{\sigma}$ orbital spreads over Cu and S ions, the antiferromagnetic moment of {\Cu} should be in between the estimates in the two extreme cases; namely, $0.22-0.45$ {\muB}/ molecule.
 The value, much smaller than the classical moment of 1 {\muB}, is understood as the spin contraction due to enhanced quantum fluctuations in low dimensions.
  
 The fitting exponent ${\beta}$ for the nuclear relaxation curve, which measures spatial distribution of {\Toner}, is shown in the inset of Fig.~\ref{Fig5}. ${\beta}$ is almost constant, around 0.9, for a wide temperature range from room temperature down to {\Tn}.
 The six protons in a molecule have different hyperfine coupling tensors against the external magnetic field, and the field is directed randomly against microcrystal axes in a powdered sample.
 This explains the small but finite deviation of ${\beta}$ from unity even above {\Tn}.
 A sudden decrease in ${\beta}$ below 13 K is a clear signature of additional inhomogeneous local field generated by the magnetic ordering.

 The nuclear spin-lattice relaxation rate, {\Toner}, above {\Tn} gives information on the spin dynamics. The room temperature value is $11-13$ s$^{-1}$, which is four times larger than the value of the analogous single-component molecular conductor {\Au} exhibiting a magnetic transition at 110 K.
 The {\Au} salt remains metallic even below {\Tn}; so the origin of the magnetic phase transition has been argued in the context of the imperfect nesting of Fermi surfaces, which is suggested by band-structure calculations.\cite{2}
 The {\Toner} is proportional to the square of fluctuations of local magnetic fields at nuclear sites. As the spin fluctuations in localized-spin systems are in general more enhanced than in the degenerate metallic state, it is reasonable that {\Toner} of {\Cu} is larger than that of {\Au}.

 In the limit of $T>>J/k_{\rm{B}}$ ($k_{\rm{B}}$ : Boltzmann constant),\cite{10} {\Toner} of a powdered sample with localized spins is given as
\begin{equation}\label{eq1}
T_{1}^{-1}=\frac{\sqrt{8{\pi}}g^{2}{\gamma}_{\rm N}^{2}}{3{\omega}_{\rm e}}S(S+1) \sum_{i}B_{i}^{2},
\end{equation}
where
\begin{equation*}
{\omega}_{\rm e}^{2}=\frac{8zJ^{2}S(S+1)}{3{\hbar}^{2}}.
\end{equation*}
Here $g$ is electron g-factor, ${\gamma}_{\rm N}$ is the gyromagnetic ratio of the {\proton} nuclei, $J$ is the exchange interaction between spins of $S$, $z$ is the coordination number, and $B_{i}^{2}$ is the solid-angle average of the dipole hyperfine coupling constant between the {\proton} nuclear spin and the Cu spin labeled by $i$($i=1-6$).
 If the $dp{\sigma}$ spins are solely on the Cu ions, $B_{i}$ is given as $B_{i}=${\muB}/$r_{i}^{3}$, where {\muB} is Bohr magneton and $r_{i}$ is a distance between proton and the $i$-th Cu spin.
 At a given proton site, dipole fields from the six nearest Cu spins are calculated and then the squared value, $B_{i}^{2}$ are summed up with respect to $i$.
 The measured relaxation rate is the averaged value over the six proton sites, $<${\Toner}$>$.
 (Note that the relaxation rate from different Cu spins should be summed up at a given proton site whereas the relaxation rates at different proton sites should be averaged.)
 The substitution of the values of $<\sum_{i} B_{i}^{2}>=7.5{\times}10^{4}$ (Oe/{\muB})$^{2}$, $g=2.048$, $J=169$ K and $z=2$ into Eq. (\ref{eq1}) yields $<${\Toner}$>$ of 6.4 s$^{-1}$.
 If the spins are on four S atoms around Cu, $<\sum_{i}B_{i}^{2}>$ is estimated at $2.0{\times}10^{4}$ (Oe/{\muB})$^{2}$, which yields {\Toner} of 1.7 s$^{-1}$.
 In reality, the $dp{\sigma}$ spins are distributed on Cu and S atoms and therefore the high-temperature limit of $<${\Toner}$>$ is expected to be in between them.
 This value gives a reasonable explanation to the experimental value of 12 s$^{-1}$, considering an ambiguity in estimating $B_{i}$ and averaging the calculated values over the non-equivalent proton sites, and a possible contribution to $<${\Toner}$>$ from the isotropic hyperfine coupling through the bonding electrons within a molecule.

 The {\Toner} is a measure of the dynamic part of spin susceptibility and its temperature dependence is often useful to know the nature of the spin system in question.
 Generally, low-dimensional spin systems do not show a magnetic ordering even at $T{\sim}J/k_{\rm B}$.
 The transition, if any, occurs at a lower temperature far below $J/k_{\rm B}$ as in the present case.
 {\Cu} is regarded as a quasi-one-dimensional spin system with $J$ of 169 K.
 In the just vicinity of {\Tn} of 13 K, {\Toner} shows a steep variation due to the critical fluctuations;
 however, the temperature dependence is featured by the temperature-insensitivity for 20 K$<T<$40 K, and a crossover to a temperature-linear variation for 60 K$<T<$160 K (Fig.~\ref{Fig5}).
 This behavior is consistent with the scaling theory for the $S=1/2$ Heisenberg chains model.\cite{11} It predicts that the staggered susceptibility, which gives a constant {\Toner}, dominates the low-temperature behavior at $T<<J/k_{\rm B}$ whereas at higher temperature (but below $J/k_{\rm B}$), {\Toner} is dominated by uniform susceptibility, leading to temperature-linear behavior.
 Quantum Monte Carlo calculations also suggest the characteristic crossover between the two behaviors (Fig. 3 in Ref. 12). The predicted low-temperature regime is seen in the temperature-insensitive Cu-NMR {\Toner} in a conventional one-dimensional system Sr$_{2}$CuO$_{3}$ with $J=2200{\pm}200$ K and {\Tn}$=5$ K,\cite{13, 14}
 and the overall crossover behavior is seen in $^{31}$P-NMR {\Toner} of another one-dimensional spin system BaCuP$_{2}$O$_{7}$ with $J=108{\pm}3$ K and {\Tn}$=0.85$ K;\cite{15}
 that is, {\Toner} is nearly temperature-independent below $k_{\rm B}T/J{\sim}0.12$ and linear with temperature above the temperature quite similar to the present results.
 According to Ref. 12, the crossover temperature depends on the non-locality of the form factor in the hyperfine coupling constant.
 Anyway, the data in Fig.~\ref{Fig5} corroborate that the spins form a one-dimensional system as suggested earlier.

 As described in Introduction, molecular orbital calculations indicate that the $p{\pi}$ and $dp{\sigma}$ orbitals are energetically close in the present system.
 This reserves another possibility that the localized spins are on the $p{\pi}$ orbital. In what follows, we argue this case.
 In an isolated Cu(tmdt)$_{2}$ molecule, the (symmetric or asymmetric) $p{\pi}$ orbital is constructed with a weak coupling (bonding or anti-bonding) of the two tmdt $p{\pi}$ orbitals on both sides of Cu.
 In the crystal, the tmdt $p{\pi}$ orbitals have overlapping with those in adjacent molecules as well. Referring to first-principles calculations and its reduction to tight-biding model on the isostructural {\Au},\cite{16} the largest inter-molecular transfer integral between the tmdt ligands is 208 meV, which is four times of the intra-molecular one (54 meV).
 This means that an entity to accommodate a spin is the inter-molecular (tmdt)$_{2}$ instead of the intra-molecular (tmdt)$_{2}$. Taking other transfer integrals available in Ref. 16 into account, the spins on the (tmdt)$_{2}$ would form an anisotropic three-dimensional lattice with transfer integrals of 50 meV in $ab$ plane and of its half in $c$ direction.
 A qualitatively similar situation is expected in {\Cu}. If spins were in tmdt ligands, one would have three-dimensional spin dynamics.
 This consequence is inconsistent with the one-dimensional nature of the spin dynamics in the present observation.

 Therefore, the observed behavior of {\Toner} gives an experimental support for the picture that electron spins are on the $dp{\sigma}$ orbital in {\Cu}.

\section{CONCLUSION}
 We have performed {\proton}-NMR experiments for the single-component molecular system, {\Cu}.
 The observation of a NMR line broadening and a divergent peak against temperature provides clear microscopic evidences for an antiferromagnetic transition at 13 K.
 The analysis of the broadened spectra below the transition temperature gives a moment of $0.22-0.45$ {\muB}/molecule.
 The temperature dependence of NMR relaxation rate above the transition temperature points to spin dynamics of one-dimensional nature, which strongly suggests that spins are on $dp{\sigma}$ orbital in {\Cu} unlike other members in the {\Mtmdt}.
 This consequence shows a potential richness in electronic phases in {\Mtmdt}.

\section{ACKNOWLEDGEMENTS}
 The authors thank H. Seo, S. Ishibashi and H. Fukuyama for valuable discussions.
 This work was supported by MEXT Grant-in-Aids for Scientific Research on Innovative Area (New Frontier of Materials Science Opened by Molecular Degrees of Freedom; Grants No. 20110002 and No. 20110003),
 JSPS Grant-in-Aids for Scientific Research (A) (Grant No. 20244055), (C) (Grant No. 20540346), and (B) (Grant No. 20350069),
 and MEXT Global COE Program at University of Tokyo (Global Center of Excellence for the Physical Sciences Frontier; Grant No.G04).

\begin{thebibliography}{99}
 \bibitem{1}A. Kobayashi, Y. Okano, and H. Kobayashi,\ J.\ Phys.\ Soc.\ Jpn.\ {\bf 75},\ 051002\ (2006).
 \bibitem{2}S.\ Ishibashi,\ K.\ Terakura,\ and\ A.\ Kobayashi,\ J.\ Phys.\ Soc.\ Jpn.\ {\bf 77},\ 024702\ (2008).
 \bibitem{3}S.\ Ishibashi\ and K.\ Terakura,\ JPS2010\ spring\ meeting.
 \bibitem{4}H.\ Tanaka,\ M.\ Tokumoto,\ S.\ Ishibashi,\ D.\ Graf,\ E.\ S.\ Choi,\ J.\ S.\ Brooks,\ S.\ Yasuzuka,\ Y.\ Okano,\ H.\ Kobayashi,\ and\ A.\ Kobayashi,\ J.\ Am.\ Chem.\ Soc.\ {\bf 126},\ 10518\ (2004).
 \bibitem{5}B.\ Zhou,\ M.\ Shimamura,\ E.\ Fujiwara,\ A.\ Kobayashi,\ T.\ Higashi,\ E.\ Nishibori,\ M.\ Sakata,\ H.\ Cui,\ K.\ Takahashi,\ and\ H.\ Kobayashi,\ J.\ Am.\ Chem.\ Soc.\ {\bf 128},\ 3872\ (2006).
 \bibitem{6}Y.\ Hara,\ K.\ Miyagawa,\ K.\ Kanoda,\ M.\ Shimamura,\ B.\ Zhou,\ A.\ Kobayashi,\ and\ H.\ Kobayashi,\ J.\ Phys.\ Soc.\ Jpn.\ {\bf 77},\ 053706\ (2008).
 \bibitem{7}B.\ Zhou,\ H.\ Yajima,\ A.\ Kobayashi,\ Y.\ Okano,\ H.\ Tanaka,\ T.\ Kumashiro,\ E.\ Nishibori,\ H.\ Sawa,\ and\ H.\ Kobayashi,\ Inorg.\ Chem.\ {\bf 49},\ 6740\ (2010).
 \bibitem{8}C.\ P.\ Slichter,\ Principles\ of\ Magnetic\ Resonance\ (Springer,\ Heidelberg,\ 1998).
 \bibitem{9}A.\ Kobayashi\ (private communication).
 \bibitem{10}T.\ Moriya,\ Prog.\ Theor.\ Phys.\ {\bf 16},\ 23\ (1956);\ {\bf 16},\ 641\ (1956).
 \bibitem{11}S.\ Sachdev,\ Phys.\ Rev.\ B\ {\bf 50},\ 13006\ (1994).
 \bibitem{12}A.\ W.\ Sandvik,\ Phys.\ Rev.\ B\ {\bf 52},\ R9831\ (1995).
 \bibitem{13}N.\ Motoyama,\ H.\ Eisaki,\ S.\ Uchida,\ Phys.\ Rev.\ Lett\ {\bf 76},\ 3212\ (1996).
 \bibitem{14}M.\ Takigawa,\ N.\ Motoyama,\ H.\ Eisaki,\ and\ S.\ Uchida,\ Phys.\ Rev.\ Lett\ {\bf 76},\ 4612\ (1996). 
 \bibitem{15}R.\ Nath,\ A.\ V.\ Mahajan,\ N.\ Buttgen,\ C.\ Kegler,\ A.\ Loidl,\ and\ J.\ Bobroff,\ Phys.\ Rev.\ B\ {\bf 71},\ 174436\ (2005).
 \bibitem{16}H.\ Seo,\ S.\ Ishibashi,\ Y.\ Okano,\ H.\ Kobayashi,\ A.\ Kobayashi,\ H.\ Fukuyama,\ and\ K.\ Terakura,\ J.\ Phys.\ Soc.\ Jpn.\ {\bf 77},\ 023714\ (2008).
\end{thebibliography}
\end{document}